\documentclass[english]{llncs}
\usepackage[T1]{fontenc}
\usepackage[latin1]{inputenc}
\usepackage{float}
\usepackage{graphicx}

\makeatletter

\floatstyle{ruled}
\newfloat{algorithm}{tbp}{loa}
\floatname{algorithm}{Algorithm}

\def\draftversion{N}        
\def\note[#1]#2{\message{(#1)}\if\draftversion
{\noindent\em[#2]\/}\fi}
\if \draftversion Y
\fi

\usepackage{babel}
\makeatother
\begin{document}

\title{Identifying Logical Homogeneous Clusters for Efficient Wide-area
Communications}

\author{Luiz Angelo B. Estefanel\thanks{Supported by grant BEX 1364/00-6 from CAPES - Brazil}
and Grégory Mounié}

\institute{Laboratoire ID - IMAG, Project APACHE\thanks{This project is supported by CNRS, INPG, INRIA and UJF}
\\
51, Avenue Jean Kuntzmann, F-38330 Montbonnot St. Martin, France\\
 \email{\{Luiz-Angelo.Estefanel,Gregory.Mounie\}@imag.fr}}

\maketitle
\begin{abstract}
Recently, many works focus on the implementation of collective communication
operations adapted to wide area computational systems, like computational
Grids or global-computing. Due to the inherently heterogeneity of
such environments, most works separate {}``clusters'' in different
hierarchy levels. to better model the communication. However, in our
opinion, such works do not give enough attention to the delimitation
of such clusters, as they normally use the locality or the IP subnet
from the machines to delimit a cluster without verifying the {}``homogeneity''
of such clusters. In this paper, we describe a strategy to gather
network information from different local-area networks and to construct
{}``logical homogeneous clusters'', better suited to the performance
modelling.

\vspace{-0.1cm}
\end{abstract}

\section{\label{sec:Introduction}Introduction}

In recent years, many works focus on the implementation of collective
communications adapted to large-scale systems, like Grids. While the
initial efforts to optimise such communications just simplified the
models to assume equal point to point latencies between any two processes,
it becomes obvious that any tentative to model practical systems should
take in account the inherently heterogeneity of such systems. This
heterogeneity represents a great challenge to the prediction of communication
performance, as it may come from the distribution of processors (as
for example, in a cluster of SMP machines), from the distance between
machines and clusters (specially in the case of a computational Grid)
and even from variations in the machines performance (network cards,
disks, age of the material, etc.). It is also a true concern for users
that run parallel applications over their LANs, where there can be
combined different machines and network supports. 

As the inherent heterogeneity and the growth of computational Grids
make too complex the creation of full-customised collective operations,
as proposed in the past by \cite{key-1,key-10}, a solution followed
by many authors is to subdivide the network in communication layers.
Most systems only separate inter and intra-cluster communications,
optimising communication across wide-area networks, which are usually
slower than communication inside LANs. Some examples of this {}``two-layered''
approach include \cite{key-6,key-15,key-27,key-5}, where ECO \cite{key-6,key-27}
and MagPIe \cite{key-6} apply this concept for wide-area networks,
and LAM-MPI 7 \cite{key-15} applies it to SMP clusters. Even though,
there is no real restriction on the number of layer and, indeed, the
performance of collective communications can still be improved by
the use of multi-level communication layers, as observed by \cite{key-25,key-19}. 

While the separation of the network in different levels can improve
the communication performance, it still needs to be well tuned to
achieve optimal performance levels. To avoid too much complexity,
the optimisation of two-layer communication or the composition of
multiple layers relies on a good communication modelling of the network.
While in this work we use pLogP \cite{key-9}, the main concern for
the accuracy of a network model relies on the homogeneous behaviour
of each cluster. If there are some nodes that behave differently from
what was modelled, they will interfere with the undergoing operation.
It is worth to note, however, that most of the works on network-aware
collective communication seem to ignore this problem, and define clusters
according to simple {}``locality'' parameters, as for example, the
IP subnet of the nodes.

While there are many network monitoring tools that could help on the
identification of such heterogeneities like, for example, NWS \cite{key-29},
REMOS \cite{key-5} or TopoMon \cite{key-22}, they still do not provide
information about machines that hold multiple application processes,
like SMP machines. Further, these tools are unable to identify heterogeneities
due to the application environment, as for example, the use of an
IMPI \cite{key-21} server to interconnect different MPI distributions,
or an SSH tunnel among different clusters protected by a firewall.

In this paper, we describe a framework to allow the gathering of independent
network information from different clusters and the identification
of {}``logical clusters''. Our proposal combines the detection of
{}``homogeneity islands'' inside each cluster with the detection
of SMP processors, allowing the stratification of the network view,
from the node layer (specially in the case of SMP machines) to the
wide-area network.

Section \ref{sec:What-we-propose} presents our proposal for automatic
topology discovery. The framework is divided in two phases. The first
one, presented on Section \ref{sec:First-Phase:-Gathering}, explains
how connectivity data collected by different clusters can be put together.
Section \ref{sec:Second-Phase:-Network} presents the second phase,
which explains how {}``logical clusters'' are defined from the collected
data, and how SMP nodes can be identified. Section \ref{sec:Practical Results}
presents the results from a practical experiment, and some considerations
on the benefits from the use of our framework. Finally, Section \ref{sec:Conclusions}
presents our conclusions and perspective for future works.

\vspace{-0.1cm}

\section{\label{sec:What-we-propose}What we propose}

We propose a method to automatically discover network topology in
order to allow the construction of optimised multilevel collective
operations. We prefer automatic topology discovery instead of a predefined
topology because if there are hidden heterogeneities inside a cluster,
they may interfere with the communication and induce a non negligible
imprecision in the communication models. 

The automatic discovery we propose should be done in two phases: the
first phase collects reachability data from different networks. The
second phase, executed at the application startup, identifies SMP
nodes (or processes in the same machine), subdivides the networks
in homogeneous clusters and acquires pLogP parameters to model collective
communications. 

As the first step is independent from the application, it can use
information from different monitoring services, which are used to
construct a distance matrix. This distance matrix does not need to
be complete, in the sense that a cluster does not need to monitor
its interconnection with other clusters, and several connectivity
parameters can be used to classify the links and the nodes as, for
example, latency and throughput. 

When the network is subdivided in homogeneous subnets, we can acquire
pLogP parameters, necessary to model the collective communications
and to determine the best communication schedule or hierarchy. Due
to the homogeneity inside each subnet, pLogP parameters can be obtained
in an efficient way, which reflects in a small impact on the application
initialisation time. 

At the end of this process we have logical clusters of homogeneous
machines and accurate interconnection parameters, that can be used
to construct an interconnection tree (communicators and sub-communicators)
that optimises both inter and intra-cluster communication.\vspace{-0.1cm}

\section{\label{sec:First-Phase:-Gathering}First Phase: Gathering Network
Information}

While there are many works that focus on the optimisation of collective
communications in Grid environments, they consider for simplicity
that a cluster is defined by its locality or IP subnet, and that all
machines inside a cluster behave similarly. Unfortunately, this {}``locality''
assumption is not adequate to real systems, which may contain machines
that behave differently both in performance and in communication.
In fact, even in clusters with similar material, machines can behave
differently (we believe that it is nearly impossible to have homogeneity
in a cluster with hundreds of machines). Hence, to better optimise
collective communications in a Grid environment, the choice of the
topologies must be based on operational aspects that reflect the real
performance level of each machine or network.

\vspace{-0.1cm}

\subsection{\label{sub:Obtaining-Network-Metrics}Obtaining Network Metrics}

There are many tools specialised on network monitoring. These tools
can obtain interconnectivity data from direct probing, like for example
NWS \cite{key-29}, from SNMP queries to network equipments, like
REMOS \cite{key-5}, or even combine both approaches, like TopoMon
\cite{key-22}. For simplicity, this work obtains data at the application
level, with operations built according to NWS definition. We chose
NWS as it is a \emph{de facto} standard in the Grid community, and
can be configured to provide information like communication latency,
throughput, CPU load and available memory. To our interest, we can
use communication latency and throughput, obtained from NWS, to identify
sets of machines with similar communication parameters. 

However, contrarily to some tools like TopoMon, our method does not
require total interconnection among all nodes in all clusters. Indeed,
the objective of the first step of our topology discovery is to identify
heterogeneity inside each cluster, and by this reason, each cluster
may use its own monitoring tool, without being aware of other clusters.
This strategy allows the use of regular monitoring data from each
cluster, while does not create useless traffic between different clusters.
Hence, the data obtained from different clusters is collected and
used to construct a distance matrix, which will guide the elaboration
of the cluster hierarchy for our collective operations. 

As clusters are not aware of each other, the {}``missing interconnections''
clearly delimit their boundaries, which reduces the cost of the clustering
process. Moreover, we are not strongly concerned with the problem
of shared links, like \cite{key-22} or \cite{key-30}, because the
reduction of the number of messages exchanged among different clusters
is part of the collective communication optimisation.

\vspace{-0.1cm}

\section{\label{sec:Second-Phase:-Network}Second Phase: Application-level
Clustering}

One reason for our emphasis on the construction of logical clusters
is that machines may behave differently, and the easiest way to optimise
collective communications is to group machines with similar performances.
In the following section we describe how to separate machines in different
logical clusters according to the interconnection data we obtained
in the First Phase, how to identify processes that are in the same
machine (SMP or not), and how this topology knowledge may be used
to obtain pLogP parameters in an efficient way.

\vspace{-0.1cm}

\subsection{\label{sub:Clustering}Clustering}

From the interconnection data from each cluster acquired on the previous
phase, we can separate the nodes in different {}``logical cluster''.
To execute this classification, we can use an algorithm similar to
the Algorithm \ref{cap:ECO-algorithm-for}, presented by ECO\cite{key-5}.

\begin{algorithm}

\caption{\label{cap:ECO-algorithm-for}ECO\cite{key-5} algorithm for partitioning
the network in subnets}

\texttt{\scriptsize initialize subnets to empty}{\scriptsize \par}

\texttt{\scriptsize for all nodes}{\scriptsize \par}

\texttt{\scriptsize ~~node.min\_edge = minimum cost edge incident
on node}{\scriptsize \par}

\texttt{\scriptsize sort edges by nondecreasing cost}{\scriptsize \par}

\texttt{\scriptsize for all edges (a,b)}{\scriptsize \par}

\texttt{\scriptsize ~~if a and b are in the same subnet}{\scriptsize \par}

\texttt{\scriptsize ~~~~continue}{\scriptsize \par}

\texttt{\scriptsize ~~if edge.weight>1.20 {*} node(a).min\_edge
or edge.weight>1.20 {*} node(b).min\_edge}{\scriptsize \par}

\texttt{\scriptsize ~~~~continue}{\scriptsize \par}

\texttt{\scriptsize ~~if node (a) in a subnet}{\scriptsize \par}

\texttt{\scriptsize ~~~~if (edge.weight>1.20 {*} node(a).subnet\_min\_edge)}{\scriptsize \par}

\texttt{\scriptsize ~~~~~~continue}{\scriptsize \par}

\texttt{\scriptsize ~~if node (b) in a subnet}{\scriptsize \par}

\texttt{\scriptsize ~~~~if (edge.weight>1.20 {*} node(b).subnet\_min\_edge)}{\scriptsize \par}

\texttt{\scriptsize ~~~~~~continue}{\scriptsize \par}

\texttt{\scriptsize ~~merge node(a).subnet and node(b).subnet}{\scriptsize \par}

\texttt{\scriptsize ~~set subnet\_min\_edge to min(edge,node(a).subnet\_min\_edge,
node(b).subnet\_min\_edge)}
\end{algorithm}

This algorithm analyses each interconnection on the distance matrix,
grouping nodes for wich their incident edges respect a latency bound
(20\%, by default) inside that subnet. As this algorithm does not
generate a complete hierarchy, just a list of subnets, it does not
impose any hierarchical structure that would {}``freeze'' the topology,
forbidding the construction of dynamic inter-cluster trees adapted
to each collective communication operation and its parameters (message
size, segments size, etc.). 

\vspace{-0.1cm}

\subsection{\label{sub:Identifying-SMP-nodes:}SMP Nodes and Group Communicators}

While NWS-like tools provide enough information to identify logical
clusters, they cannot provide information about processes in SMP nodes,
as they are created by the application. Actually, as the processes
distribution depends on the application parameters (and environment
initialisation), the identification of processes in SMP nodes shall
be done during the application startup.

However, the implementation of an SMP-aware MPI is not easy, because
the definition of MPI does not provides any procedure to map process
ranks into real machine names. To exemplify this difficulty, we take
as example the recent version 7 from LAM/MPI \cite{key-15}. Their
SMP aware collective communications, based on MagPIe \cite{key-6},
rely on the identification of processes that are started in the same
machine, but they use proprietary structures to identify the location
of each process. To avoid be dependent on a single MPI distribution,
we adopted a more general solution, where each process, during its
initialisation, call \emph{gethostname()}, and sends this data to
a {}``root'' process that will centralise the analysis. If perhaps
this approach is not as efficient as the one used by LAM, it still
allows the identification of processes in the same machine (what can
be assumed as an SMP machine). 

As the data received by the root contains both the machine name and
the process rank, it can translate the logical clusters into communicators
and sub-communicators, adapted to the MPI environment. 

\vspace{-0.1cm}

\section{\label{sec:Practical Results}Practical Results}

\vspace{-0.1cm}

\subsection{Clustering}

To validate our proposal, we looked for a network environment that
could raise some interesting scenarios to our work. Hence, we decided
to run our tests on our experimental cluster, IDPOT%
\footnote{http://idpot.imag.fr%
}. IDPOT can be considered as a {}``distributed cluster'', as its
nodes are all spread through our laboratory, while connected with
a dedicated Gigabit Ethernet (two switches). All machines are Bi-Xeon
2.5 GHz, with Debian Linux 2.4.26, but they have network card from
two different manufacturers, and the distribution of the machines
in the building may play an important role in the interconnection
distance between them and the Gigabit switches. 

\begin{figure}
\begin{center}\includegraphics[%
  bb=0bp 25bp 650bp 295bp,
  width=0.85\linewidth]{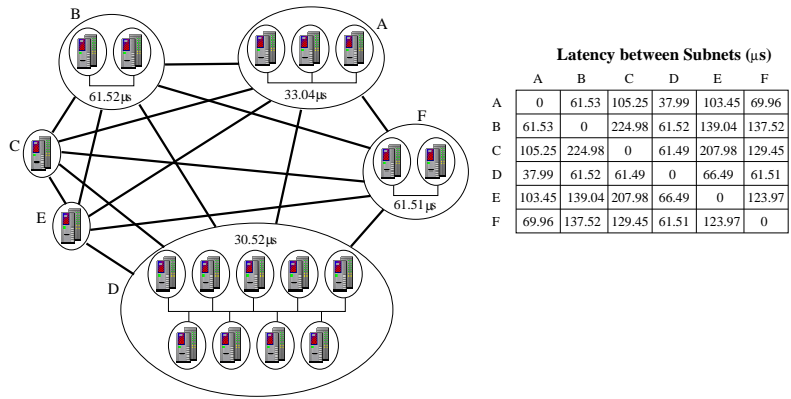}\end{center}

\caption{\label{Figure:IDPOT-network-partition}IDPOT network partition, with
latency among nodes and subnets}
\end{figure}

Applying the methods described in this paper over a group of 20 machines
from IDPOT gives the following result, depicted on Fig. \ref{Figure:IDPOT-network-partition}.
This figure presents the resulting subnets, as well as the interconnection
times between each subnet and among nodes in the same subnet. It is
interesting to note how the relative latency among each cluster would
affect a collective communication that was not aware of such information.
For example, in the case of a two-level model, subnets C and E would
affect the expected performance, as their interconnections are twice
or even three times slower than others. This would also reflect in
the case of a multi-layer model, where an unaware algorithm could
prefer to connect directly subnets C and E, while it is more interesting
to forward communications through subnet D. 

\vspace{-0.5cm}We identified as the main factor for such differences
the presence of network cards from one manufacturer on subnets A and
D, while subnets B, C, E and F have onboard cards from other manufacturer.
As second factor, we can list the location of the nodes. While it
played a less important role, the location was the main cause for
separation between subnet A and subnet D. Actually, the distance between
those machines, which are under different switches, affected the latency
just enough to force ECO's algorithm to separate them in two different
subnets. A correct tuning on the parameters from ECO's algorithm may
allow subnets A and D to be merged in a single one, a more interesting
configuration for collective communications.

\vspace{-0.1cm}

\subsection{\label{sub:Getting-pLogP-data}Efficient Acquisition of pLogP Parameters}

While the logical clusters generated by our framework allow a better
understanding of the network effective structure, we are still unable
to model communications with precision. This first reason is that
interconnection data may be incomplete. As said in Section \ref{sub:Obtaining-Network-Metrics},
the monitoring tools act locally to each LAN, and by this reason,
they do not provide data from the inter-cluster connections. 

\begin{figure}
\begin{center}\includegraphics[%
  bb=0bp 35bp 522bp 140bp,
  width=0.60\linewidth,
  keepaspectratio]{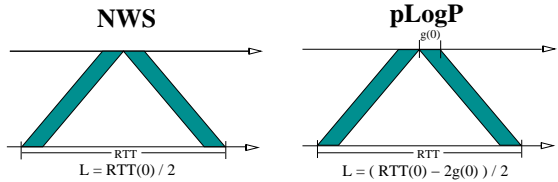}\end{center}

\caption{\label{Figure:Differences-between-NWS}Differences between NWS and
pLogP {}``latency''}
\end{figure}

Besides this, the data acquired by the monitoring tools is not the
same as the data used in our models. For example, the latency, which
originally should have the same meaning to the monitoring tool and
the application, is obtained differently by NWS and pLogP. In NWS,
the latency is obtained directly from the round-trip time, while pLogP
separates the round-trip time in latency and gap, as depicted by Figure
\ref{Figure:Differences-between-NWS}, with differences that may interfere
on the communication model. In addition, the information collected
by the monitoring tools is not obtained by the application itself,
and thus, is not submitted to the same constraints that the application
will find at runtime, as for example, the use of an Interoperable
MPI (IMPI) server to interconnect the clusters. 

\vspace{-0.5cm}Hence, to model the communication in our network,
we need to obtain parameters specifically for pLogP. Hopefully, there
is no need to execute $n(n-1)$ pLogP measures, one for each possible
interconnection. The first reason is that processes belonging to the
same machine were already identified as SMP processes and grouped
in specific sub-communicators. And second, the subnets are relatively
homogeneous, and thus, we can get pLogP parameters in an efficient
way by considering a single measure inside each subnet as a sample
from the pLogP parameters common to the entire cluster. As one single
measure may represents the entire subnet, the total number of pLogP
measures is fairly reduced. If we sum up the measures to obtain the
parameters for the inter-clusters connections, we shall execute at
most C(C+1) experiments, where C means the number of subnets. Further,
if we suppose symmetrical links, we can reduce this number of measures
by half, as $a\rightarrow b=b\rightarrow a$. By consequence, the
acquisition of pLogP parameters for our experimental 20-machines cluster
would need at most $(6*(6+1))/2=21$ measures.

\vspace{-0.1cm}

\section{\label{sec:Conclusions}Conclusions}

This paper proposes a simple and efficient strategy to identify communication
homogeneities inside computational clusters. The presence of communication
heterogeneities reduces the accuracy from the communication models
used to optimise collective communications in wide-area networks.
We propose a low cost method that gathers connectivity information
from independent clusters and groups nodes with similar characteristics.
Using real experiments on one of our clusters, we show that even minor
differences may have a direct impact on the communication performance.
Our framework allowed us to identify such differences, classifying
nodes accordingly to their effective performance. Using such classification,
we can ensure a better accuracy for the communication models, allowing
the improvement of collective communication performances, specially
those structured on multiple layers.

\end{document}